# Fast quantum key distribution with decoy number states

D. ACHILLES, E. ROGACHEVA, and A. TRIFONOV
MagiQ Technologies, Inc., 11 Ward Street, Somerville, MA 02143
Email: d.achilles@magiqtech.com

**Abstract:** We investigate the use of photon number states to identify eavesdropping attacks on quantum key distribution (QKD) schemes. The technique is based on the fact that different photon numbers traverse a channel with different transmittivity. We then describe two QKD schemes that utilize this method, one of which overcomes the upper limit on the key generation rate imposed by the dead time of detectors when using a heralded source of photons.

## 1. Introduction

Quantum key distribution (QKD) is a method of generating a secret key between two parties (Alice, the sender, and Bob, the receiver) that is provably secure assuming that the laws of quantum physics are correct and that any eavesdropper (Eve) must work within the framework of these laws. Introduced in 1984, the BB84 protocol for QKD is based on the use of single photons for encoding the quantum information [1]. The uncertainty principle guarantees the security of the protocol since Eve does not have any *a priori* information about the basis that Alice used for encoding the information, ensuring the impossibility of a precise measurement of the secure bit due to the Heisenberg uncertainty principle. Sometimes the security is formulated in terms of the no-cloning theorem by Wooters and Zurek [2], which asserts that it is impossible for Eve to generate a precise copy of the unknown photonic information being sent from Alice to Bob without destroying said information.

The situation changes drastically if Alice does not send single photon states, but sometimes sends pulses containing two or more photons. Since Eve can now split the photons *without* destroying the information encoded in the initial state, the security is no longer protected by laws of quantum mechanics. This fact was acknowledged by the fathers of quantum cryptography [3] and has spurred long-lasting discussions on practical QKD security in the research community [4-10].

In most of the experimental realizations of the BB84 protocol weak coherent states (attenuated laser pulses) are used as the light source thus making a certain amount of multiphoton pulses available for Eve. There is a trade off in the pulse intensity between security (low mean photon number) and key generation rate (large mean photon number). But even when using so-called "true" single-photon sources, such as those based on a heralded single photon source (HSPS) using parametric downconversion (PDC) [11] or single photon emitters such as quantum dots [12] and diamond color centers [13], any practical device will produce multiple photon pulses with some non-zero probability. If the channel loss between Alice and Bob is relatively small, then Eve's attack on multiple photon pulses can be easily rejected by lowering the mean photon number and re-examining the privacy amplification model. The presence of multiple photons will result in information leakage, but this leakage can be leveraged.

The presence of single photons gives enough protection to distill an unconditionally secure key. A simple rule of thumb is: Alice and Bob must ensure that a certain amount of clicks at Bob's side result from single photon pulses sent by Alice.

If in turn the channel loss is high, and this is the most interesting situation from a practical point of view, in addition to photon splitting, Eve can take advantage of the channel loss by promoting the quantum states originating from the initial pulses containing two and more photons and suppressing the original single photon pulses [5]. We refer to this attack as the photon number splitting (PNS) attack with blocking and boosting. Lowering the mean photon number per pulse is no longer a good strategy for Alice and Bob since: 1) the key generation rate will be significantly reduced and 2) Bob's detector noise may increase the quantum bit error rate up to the level where secure key distillation is no longer possible. Let us stress again that the crucial component of the PNS attack is Eve's ability to use a loss-free channel to boost the channel transmittivity for the quantum states that she can efficiently eavesdrop, while blocking the single photon pulses.

To protect against a PNS attack and maximize both the key generation rates and the possible distance over which BB84 can be used, we must supplement the standard protocol with additional security measures to protect the key. One such method is the introduction of "decoy states" into the channel, which are used to detect an eavesdropper attempting to implement a block and boost PNS attack[14-18]. In the original decoy state method by Hwang, the decoys are weak coherent pulses which are equivalent to the pulses used to send the key in all aspects except the mean number of photons. Since Eve can not distinguish between decoy pulses and real QKD pulses, she attacks all pulses; Alice and Bob can then identify a PNS attack by comparing the gain and quantum bit error rate (QBER) for the different subsets of decoy states. Note that the Ekert protocol is immune to the PNS attack since Bell's inequality is sensitive to the presence of additional photon pairs [19-22]. For this reason utilizing remote state preparation [23, 24] (sometimes referred to as passive state preparation [25]) does not require decoys to identify an eavesdropper because performing a Bell's inequality test can reveal a PNS attack.

In the same way that single photon states can be conditionally prepared from PDC, higher order number states can also be prepared [26-28]. This requires the use of photon-number resolving detectors, which have recently become available [29-32]. Very recently the decoy state concept has been adapted to utilize multiphoton states to detect an eavesdropper [33, 34]. This manuscript introduces two practical schemes that use these multiphoton states as decoys. The first utilizes a HSPS for both the decoy and key distribution pulses. The second overcomes the speed limitations of the first by using a hybrid scheme of weak coherent pulses for key distribution and conditionally prepared number states as decoys.

## 2. Generating photon number states
In our QKD schemes, Alice and Bob will use decoy states that are one- and two-photon number states, which can be prepared by conditional state preparation of PDC with a photon-number-resolving detector (PNRD) [26-28]. Alice uses PDC to generate a two-mode squeezed state, described by:

$$|\psi_{PDC}\rangle = \sqrt{1-|\lambda|^2} \sum_{n=0}^{\infty} \lambda^n |n\rangle_A |n\rangle_B, \qquad (1)$$

where the two modes are labeled as Alice's photon (A) and Bob's photon (B), and $\lambda$ is the parametric gain of the two-mode squeezed state. She then detects her mode with a PNRD with quantum efficiency $\eta_a$, which is described by the POVM elements:

$$\hat{\Pi}_k = \sum_{n=k}^{\infty} B_k^n(\eta_a) |n\rangle\langle n|, \qquad (2)$$

where the coefficients are given by the binomial distribution

$$B_k^n(\eta_a) = \binom{n}{k} \eta_a^k (1-\eta_a)^{n-k}, \qquad (3)$$

$n$ is the number of photons in the pulse, and $k$ is the number of photons registered by Alice's PNRD. This POVM can be written in terms of *measurement operators*, or Kraus operators, as [35] $\hat{\Pi}_k = \hat{M}_k^\dagger \hat{M}_k$ where

$$M_k = \sum_{n=k}^{\infty} \sqrt{B_k^n(\eta_a)} |n\rangle\langle n|. \qquad (4)$$

The simplicity of the relation between the Kraus operators and the POVM elements is due to each term of the sum being the form of a projector. The state of the light after the measurement of Alice's mode is then determined by

$$|\psi'\rangle = \frac{\hat{M}_k |\psi_{PDC}\rangle}{\sqrt{\langle \psi_{PDC} | \hat{\Pi}_k | \psi_{PDC} \rangle}}, \qquad (5)$$

which gives the photon probability distribution for Bob's mode conditionally prepared by the registering of $k$ photons by Alice's PNRD:

$$P(n|k) = |\lambda|^{2(n-k)} \left(1-(1-\eta_a)|\lambda|^2\right)^{k+1} \binom{n}{k} (1-\eta_a)^{n-k}. \qquad (6)$$

Note that as $\eta_a \to 1$ the probability distribution becomes the delta function $P(n|k) = \delta_{n,k}$, *i.e.* Bob is sent exactly the same photon number as Alice measured [36]. The same is true when the parametric interaction is kept weak ($|\lambda|^2 \to 0$). The situation when Alice detects a single photon is demonstrated in Fig. 1. If Alice's detector efficiency is high, then she will generate photon number states with high fidelity, even with strong parametric gain, as shown in Fig. 1(a). If $\eta_a$ is low, however, she will need to be cautious of how powerful the PDC interaction is and will have to use a low parametric gain to ensure that she generates photon number eigenstates with high fidelity. Fig. 1(b) shows the case where the efficiency is low, but the parametric gain has been lowered to ensure the conditional state is approximately a photon number state. The conditionally prepared state no longer emulates a photon number state when the parametric gain remains high and the efficiency is low, as demonstrated in Fig. 1(c). Qualitatively similar results occur when Alice detects two photons ($k=2$) with her PNRD, except in this case there is no chance for a pulse to contain either zero or one.

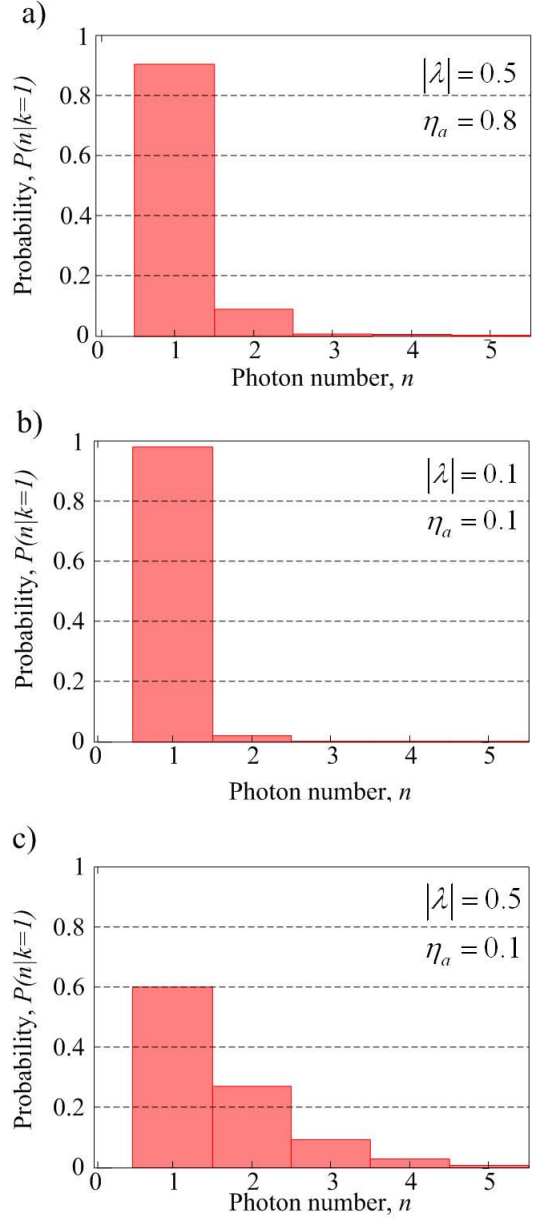

**Figure 1.** Possible conditional photon number distributions prepared by Alice as determined by Eqn. (6) for different values of Alice's detector efficiency and the parametric gain. (a) With a high detector efficiency ($\eta_a = 0.8$), the parametric gain can be large ($|\lambda| = 0.5$) and the photon number distribution is close to a photon number state. (b) When the detector efficiency is lower ($\eta_a = 0.1$), the parametric gain must be lowered ($|\lambda| = 0.1$) in order to approximate a photon number state well. (c) The photon number distribution is very different from a photon number state when the efficiency is low ($\eta_a = 0.1$) and the parametric gain is large ($|\lambda| = 0.5$). Note that for all situations there is no chance the conditionally prepared state contains zero photons.

## 3. Fock state transmittivity estimation

The state that Alice sends is, in general, a statistical mixture of Fock states:

$$\hat{\rho} = \sum_{n=0}^{\infty} p_A(n) |n\rangle\langle n|, \quad (7)$$

where $p_A(n)$ is the probability that Alice emits an $n$ photon state. We will define the transmittivity of a pulse sent by Alice as the probability that the pulse contains at least one photon when it arrives at Bob. Accordingly, each number state has a different transmittivity $\eta^{(n)}$ equal to the probability that all the photons in the pulse are not lost:

$$\eta^{(n)} = 1 - (1-\eta_c)^n, \quad (8)$$

where the channel efficiency $\eta_c$ is defined as the probability that a single photon arrives at Bob given that Alice emitted a single photon pulse. The one- and two-photon states that will be used to estimate the channel loss have transmittivities

$$\eta^{(1)} = \eta_c \quad \text{and} \quad \eta^{(2)} = \eta_c^2 + 2\eta_c(1-\eta_c) \approx 2\eta_c \quad (9)$$

where the approximation is valid for high channel loss ($\eta_c \ll 1$).

If Bob also used a PNRD they could overcome an eavesdropper by other methods [37]. However, we will assume that Bob is restricted to using binary detectors, such as avalanche photodiodes operated in Geiger-mode, which only detect the presence or absence of light and cannot resolve photon number. Such detectors will be called single-photon detectors (SPD). Alice knows when she has conditionally prepared one or two photons with her PNRD. Bob will, using classical communication, announce to Alice for which pulses he detected a click with his SPD. After a sufficient amount of data has been taken, Alice can use Bob's clicking rates to estimate the channel efficiency $\eta_c$ from the transmittivities $\eta^{(k)}$. She will have two different estimates: one based on when she detected one photon and the other from when she detected two photons.

If the estimates of $\eta_c$ give the same result, then it is safe to assume that no PNS attack is being performed by Eve. If a PNS attack were being used, the transmittivities for one and two photons would give different results for the channel efficiency. Recall that the PNS attack requires Eve to suppress the single photon pulses and "promote" the transmittivity of the two photon pulses in order for her attack to work. When she observes a two photon state, she splits off one of the photons for herself (which she will store in her quantum memory and measure later) and sends the other through a lossless channel to Bob. Therefore, for a typical PNS attack, the single photon suppression will be high $(1-\kappa_1 \approx 1)$, where $\kappa_1$ is the single photon transmittivity when Eve is performing an attack, and the two photon suppression will be low $(\kappa_2 \approx 1)$, where $\kappa_2$ is the two photon transmittivity when Eve is attacking. Since $\kappa_1 \ll \kappa_2$ the transmittivities that Alice and Bob observe will obey the relation $\eta^{(1)} \ll \eta^{(2)}$. Whereas, if the suppression is being caused by large channel loss the transmittivities obey the relation $\eta^{(2)} = 2\eta^{(1)}$.

## 4. Possible QKD implementations
There are several possibilities for how this technique may be used in QKD. We present two of them here.

### *4.1. QKD with conditionally prepared PDC*
Most implementations of QKD utilize weak coherent pulses (WCP), which have photon number distributions that obey the Poisson distribution, because they are easily generated by attenuating a laser pulse. However, one can minimize the possibility of eavesdropping (minimize multiphoton content) and increase the key distribution rate (decrease the vacuum contribution) by using a HSPS. In the basic BB84 protocol without decoy states, if Alice knew which pulses had multiple photons, as she does in the case where she has a PNRD, these pulses would be blocked or simply labeled as unusable for sifting the secure key [36]. Using the decoy state method these pulses can be utilized to detect the presence of an eavesdropper. This idea was proposed recently as a passive method of performing decoy state QKD [33] with a HSPS.

The basic scheme is shown in Fig. 2. We ignore how the information is encoded in the state preparation stage: the physical qubit could be encoded in polarization or timing. Alice pumps PDC in a nonlinear crystal (NLC) with her laser and detects her mode with a PNRD. Each time Alice detects light with her PNRD, she electronically tags the pulse with the number of photons. After Bob's detection stage, which uses SPDs he will send Alice information about which pulses he detected. From this Alice can look for an eavesdropper as discussed above. The timing information is managed via synchronization laser (Sync) and classical communication for error correction and privacy amplification occurs between the control units. The different signals can be managed, for example, by wavelength division multiplexers (WDMs) on both sides.

This scheme has several advantages over original decoy state QKD. First, it utilizes a HSPS instead of WCP for distributing the key. This allows Alice and Bob to know exactly which pulses have single photons and thus which can be used to distill the key. Second, the decoy state encoding is passive. Whereas the original decoy state protocol requires well-calibrated intensity modulators to generate pulses with different mean photon number, the different decoy states in this scheme are prepared passively by measuring the conjugate arm of PDC with a PNRD, which does not need to be precisely calibrated to prepare different photon number states. Also, the original protocol requires Bob to have a well-calibrated SPD; with our method Bob's detector does not significantly affect the one- and two-photon transmittivity estimates.

There is one major disadvantage to this technique: low pulse repetition rates, or *clock rates*. Due to the nature of the HSPS, the clock rate is limited by the deadtime and jitter of Alice's detector. For commercial SPDs, the dead-time is approximately 50ns and the jitter as low as 300ps, but for PNRD the dead-time is often much larger and is on the order of microseconds [30, 38]. Even with a 50ns dead-time, the clock rate is limited to 20MHz. On the other hand, QKD schemes that use WCP have been experimentally demonstrated with clock rates up to 10GHz [39] – nearly three orders of magnitude faster. In the following section we describe a method to get the best of both schemes: fast clock rates with photon number decoy states.

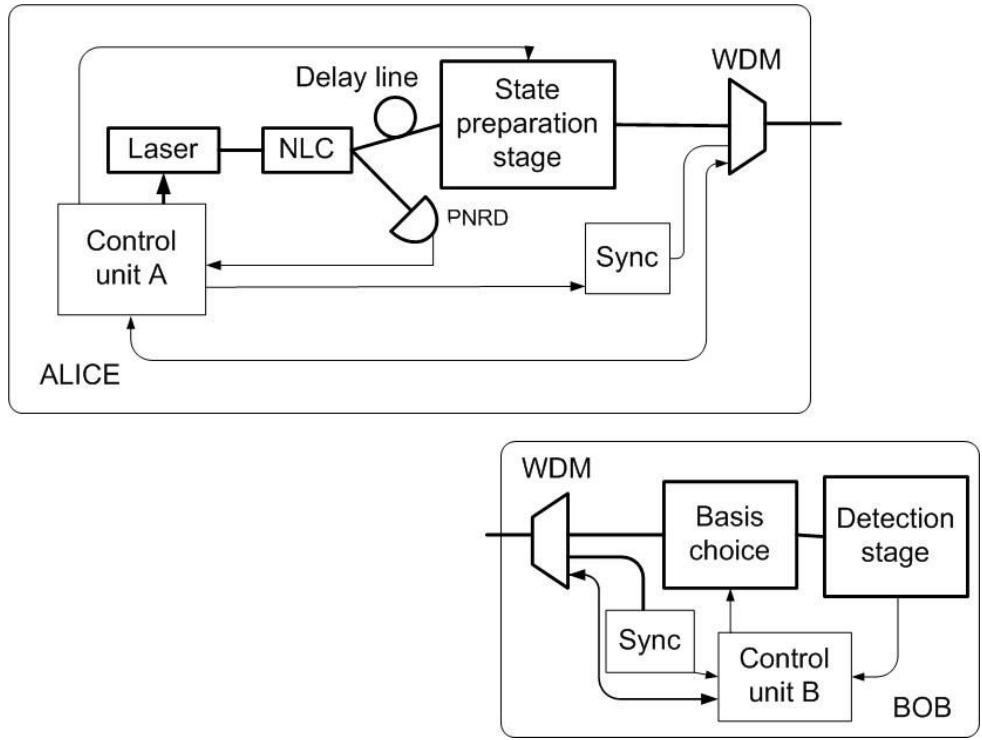

**Figure 2.** The schematic of the QKD setup using HSPS with multiphoton decoy states. Alice conditionally prepares PDC with a PNRD and encodes the key information onto the mode sent to Bob in the state preparation stage. A sync laser is used to send timing information to Bob. Classical communication also occurs between Alice and Bob's control units. The different signals are managed with a WDM on both sides. Bob performs a basis choice and detects the light with an SPD. The timing information is measured with a sync detector.

### 4.2. Fast QKD with number state decoys

The schematic of the setup is shown in Fig. 3. Due to the nature of having to combine two different sources, there are many intricacies that must be cared for in order to ensure secure QKD. The actual key distribution is done with WCP in order to ensure high clock rates. We use conditionally prepared PDC to generate photon number states which are used to detect an eavesdropper. Alice decides whether to send a WCP or a decoy from PDC using an electro-optic switch (switch 3). Since we are using different sources for the decoy states and the WCP, we must ensure that the two are completely indistinguishable in all degrees of freedom so that Eve can not selectively attack the WCP. This is achieved with a calibration step that can be run periodically by Alice during QKD.

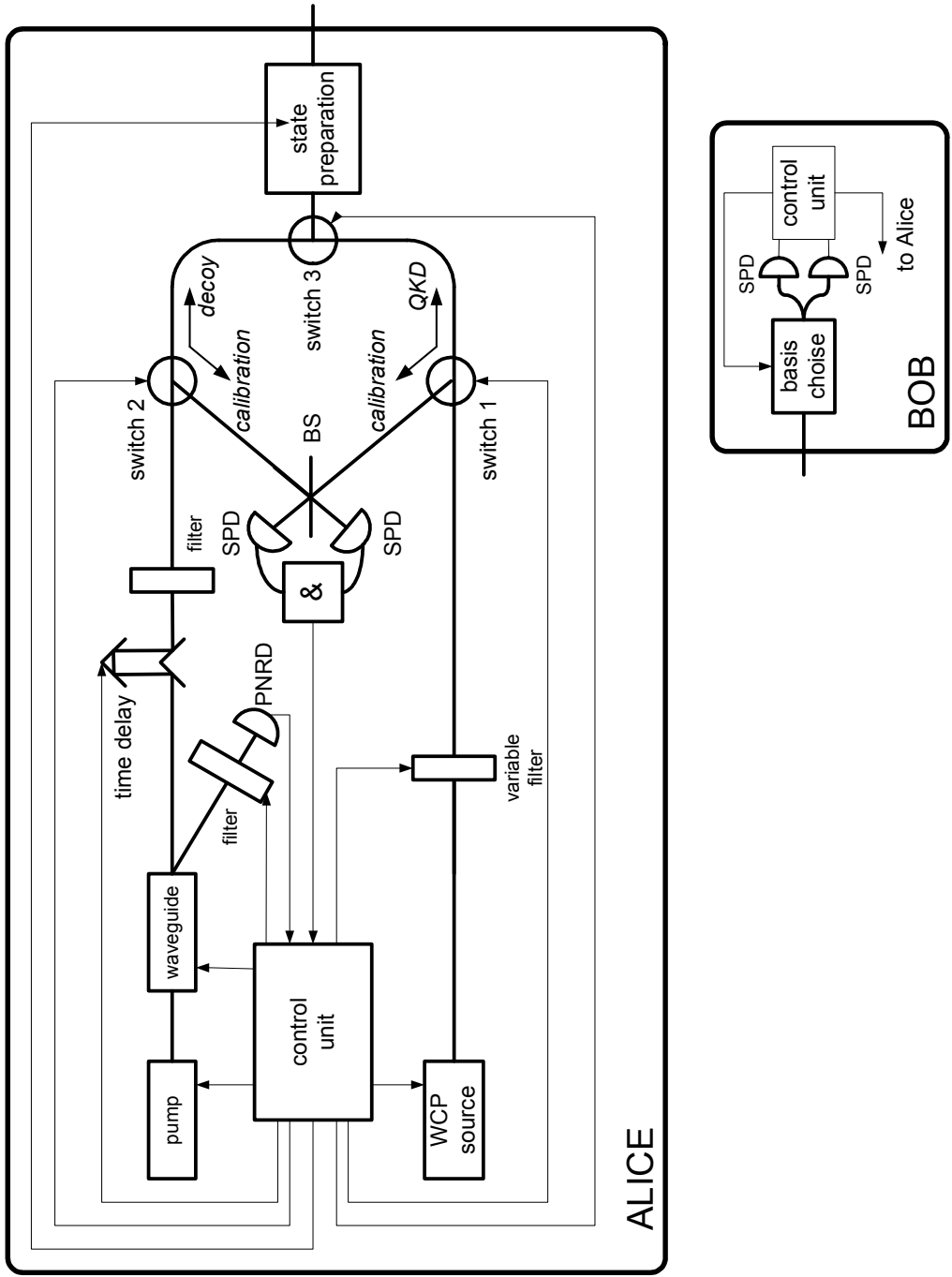

**Figure 3.** The setup for fast clock rate QKD using WCP for distributing the key and conditionally prepared PDC as decoy states. Alice can calibrate the two sources by using electro-optic switches 1 and 2 to interfere the two fields at a beam splitter (BS) with SPDs at the output to measure coincidence counts (&). Otherwise, she chooses whether to send a decoy or a WCP using switch 3. Based on the interference calibration, she can adjust the time delay and spectrum of the WCP so that the two are exactly the same for both the decoy pulse and the WCP. Bob simply makes a random basis choice and detects with SPDs. All classical communication is handled by the control units.

**4.2.1. Indistinguishability calibration.** The two crucial degrees of freedom that must be indistinguishable in the two sources are timing and spectrum. The best way to calibrate the sources is via interference because a prerequisite for any interference phenomenon is that the different possibilities that lead to a given outcome must be indistinguishable [40]. Since we wish to calibrate two different sources, we will use fourth-order interference in the form of a Hong-Ou-Mandel interferometer (HOMI) [41]. There are two parameters we will adjust to maximize interference: the time delay in the HSPS arm and the variable spectral filter in the WCP arm. To switch to calibration mode, Alice uses electro-optic switches 1 and 2 to direct the two modes towards the 50/50 BS. The two-output ports are monitored with SPDs and the coincidence counts are measured. We wish to minimize the coincidence counts, because the fourth-order interference causes photon bunching at the output of the BS. Perfect interference (100% visibility) will result in zero coincidence counts in the SPDs.

For the moment, let us assume that the two modes are both pure single photon wavepackets as described by

$$|\psi\rangle_i = \int d\omega f_i(\omega) a_i^\dagger(\omega)|0\rangle, \qquad (10)$$

where $a_i^\dagger(\omega)$ is the creation operator for a photon in mode $i$ with frequency $\omega$ and $|f_i(\omega)|^2$ is the spectrum of the single photon. Then the normalized coincidence rate of the detectors $R_c$, assuming Gaussian spectral amplitudes $f_i(\omega)$, is [42]

$$R_c = \frac{1}{2} - \frac{\sigma_1 \sigma_2}{\sigma_1^2 + \sigma_2^2} \exp\left(-\frac{(\sigma_1 \sigma_2 t)^2 + 4(\omega_i - \omega_s)^2}{2(\sigma_1^2 + \sigma_2^2)}\right), \qquad (11)$$

where $t$ is the difference in time when the two photons interact with the BS, $\sigma_i$ is the bandwidth of the $i$th field, and $\omega_i$ is the central frequency of the $i$th field. The visibility is defined as

$$V = \frac{\max(R_c) - \min(R_c)}{\max(R_c)} \qquad (12)$$

and is clearly maximum when $t = 0$, $\omega_1 = \omega_2$, and $\sigma_1 = \sigma_2$. Therefore, during the calibration, Alice will adjust the time-delay and spectrum to minimize the coincidence counts and maximize the visibility. For example, assuming the central frequencies of the two spectra have already been equalized, the visibility as a function of the ratio of bandwidths $s = \sigma_1 / \sigma_2$, as determined by Eqn. (11) and Eqn. (12) is

$$V = \frac{2s}{1 + s^2}, \qquad (13)$$

which is shown in Fig. 4.

In order to perform this calibration, it was assumed that each field was a pure single photon state. There are a couple of subtleties that need to be addressed in order to ensure successful calibration. One of the fields used is a WCP, which is not a single photon, but by maintaining a low mean photon number the visibility can be kept high [43]. The nice feature of this calibration is that even if the visibility decreases, the maximum is still located at the same place in parameter space; therefore as long as the visibility is maximized, Alice knows she has her parameters optimized.

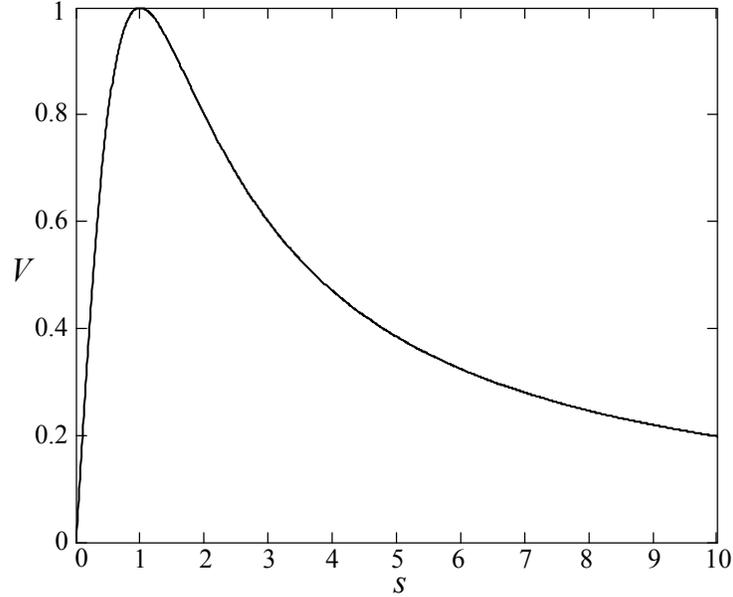

**Figure 4.** The visibility (*V*) of the HOMI (assuming equal central frequencies) as a function of the ratio between the two different bandwidths (*s*).

A second, more crucial subtlety is the assumption of purity. In order for the two fields to interfere, each photon wavepacket must be a superposition of all the different frequencies, not a statistical mixture. If they were statistical mixtures, we would be attempting to interfere a photon with one spectrum with a second that most likely has a different spectrum; this would reduce the visibility significantly. In general, most conditionally prepared PDC sources create mixed states due to the frequency entanglement that exists between the two photons. However, by choosing the dispersive properties of the nonlinear material properly, one can create a two-mode squeezed state that is factorizable in frequency and therefore creates pure single photon wavepackets upon detection of one field[44-47]. This spectral engineering is of utmost importance for this QKD protocol as well as all quantum information protocols that utilize conditionally prepared PDC as single photon sources.

**4.2.2. Choice of WCP intensity.** Once Alice and Bob have collected enough calibration data the optimal choice of mean photon number $\mu$ for her WCP can be determined from the measurement of one- and two- photon transmittivities. This is done by defining a figure of merit D to maximize with respect to the mean photon number:

$$D = p_B(1) - p_B(>1), \qquad (14)$$

where $p_B(1)$ is the probability that a pulse sent by Alice containing a single photon still contains a single photon when it arrives at Bob and $p_B(>1)$ is the probability that the pulse that originally contained multiple photons is not a vacuum state when it arrives at Bob. This figure of merit is different from the gain that is most commonly maximized [48] and maximizes the robustness of Alice and Bob's QKD rather than the key generation rate. The situation for normal channel loss is shown in Fig. 5. For perfect channel efficiency, Alice can use mean photon number $\mu = 0.5$. The optimal mean photon number decreases as the channel efficiency decreases, but it does not tend towards zero; rather, it stays above 0.316 for all possible loss values. This is essentially the same scenario as if Alice were doing a simple PNS attack without blocking single photons and boosting multiple photons.

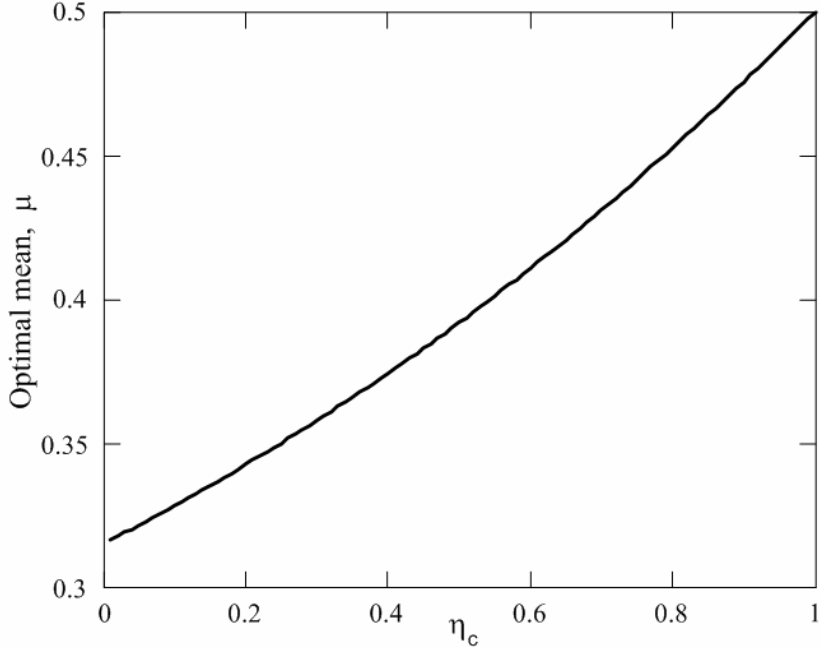

**Figure 5.** The optimal mean photon number $\mu$ for Alice's WCP as a function of the channel loss $\eta_c$. Note that even as the channel efficiency goes to zero, the optimal mean photon number remains greater than 0.316.

The situation is different when Eve is performing a PNS attack with blocking and boosting. As stated above, the key to the attack is boosting the multiphoton transmittivity $\kappa_m$ while suppressing that of a single photon $\kappa_1$. Fig. 6 shows the optimal mean photon number's dependence on the ratio $\kappa_1 / \kappa_m$ based on the same figure of merit as above. The parameter regime for an effective PNS attack is $\kappa_1 \ll \kappa_2$, resulting in a small ratio of transmittivities. This requires Alice to use a very low mean photon number for her WCP. However, if it is found that there is no PNS attack but there is simply loss (*i.e.* the parameters lie on the bold line), the mean photon number of the WCP can be much higher.

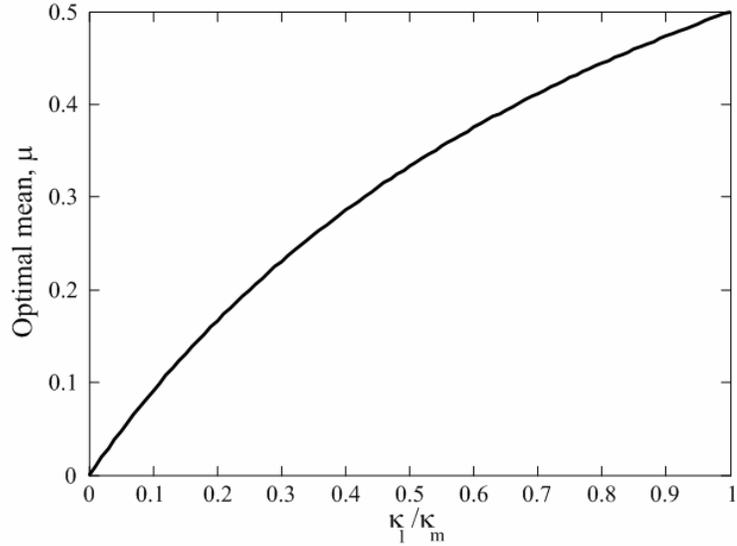

**Figure 6.** The optimal mean photon number is determined by the details of Eve's PNS attack, which is characterized by the single photon transmittivity $\kappa_1$ and the multiphoton transmittivity $\kappa_m$. The effective regime for a PNS attack is a small value for the ratio of these two numbers.

Fig. 5 is approximately incorporated in Fig. 6 because the ratio $\kappa_1 / \kappa_2 = (2 - \eta_c)^{-1}$, which ranges between 0.5 and 1 for $\eta_c = 0$ and $\eta_c = 1$, respectively. The reason the correspondence is approximate is that for the PNS attack, we assumed that the multiphoton transmittivity is the same for all photon numbers greater than one because Eve selectively boosts the transmittivity of these pulses. In the case of loss, the transmittivity is not the same for all higher photon numbers. Note that an effective PNS attack with blocking and boosting can never be confused with loss because the ratio of transmittivities can never be lower than 0.5 when it is caused by channel loss.

**5. Summary**
In conclusion, we have investigated the technique of using photon number transmittivities as a method of detecting multiphoton attacks on QKD. Two different QKD protocols based on BB84 with multiphoton decoy states were discussed as examples of this method. Though the limits on the protocols' performances were not discussed in detail, the analysis is very similar to that presented in [33]. The second of these protocols allowed the use of fast clock rates by using WCP for key distribution while using conditionally prepared photon number to detect an eavesdropper. Also, the ability to monitor the one- and two-photon transmittivities allows Alice to choose the optimal mean photon number for her WCP such that she minimizes Eve's information.